\documentclass[final,3p,twocolumn,authoryear]{elsarticle}
\usepackage{url}
\usepackage{amssymb}
\usepackage{amsmath}

\begin{document}
\begin{frontmatter}


\title{Lagrangian study of surface transport in the Kuroshio Extension area based on simulation of propagation of Fukushima-derived radionuclides}


\author{S.V. Prants}
\ead{prants@poi.dvo.ru}
\ead[url]{www.dynalab.poi.dvo.ru}

\author{M.V. Budyansky}
\author{M.Yu. Uleysky}

\address{Laboratory of Nonlinear Dynamical Systems,\\
Pacific Oceanological Institute of the Russian Academy of Sciences,\\
43 Baltiiskaya st., 690041 Vladivostok, Russia\\
URL: htpp://www.dynalab.poi.dvo.ru}

\begin{abstract}
Lagrangian approach is applied to study near-surface large-scale transport
in the Kuroshio Extension area using a simulation with synthetic particles
advected by AVISO altimetric velocity field.
A material line technique is applied  to find the origin of water
masses in cold-core cyclonic rings pinched off
from the jet in summer 2011. Tracking and Lagrangian maps provide the evidence of cross-jet transport.
Fukushima derived caesium isotopes are used as Lagrangian tracers
to study transport and mixing in the area a few months after the March of 2011 tsunami that caused a heavy damage
of the Fukushima nuclear power plant (FNPP).
Tracking maps are computed to trace the origin of
water parcels with measured levels of $^{134}$Cs and $^{137}$Cs
concentrations collected  in two R/V cruises in June and July 2011 in the large area of the Northwest Pacific
\citep{Kaeriyama13,Buesseler12}.
It is shown that Lagrangian simulation is useful to finding the
surface areas that are potentially dangerous due to the risk of radioactive
contamination.
The results of simulation are supported by tracks of the surface drifters
which were deployed in the area.
\end{abstract}
\end{frontmatter}

\section{Introduction}

The Kuroshio Extension (KE) prolongs the Kuroshio Current, a western boundary current in the Northwest Pacific,
when the latter separates from the continental shelf
of the Japanese island Honshu at Cape Inubo about $35^{\circ}42'$~N. It flows eastward from this point as a strong
unstable meandering jet constituting a front separating the warm subtropical and cold subpolar waters of the
North Pacific Ocean. There are cyclonic and anticyclonic recirculation gyres on the northern
and southern flanks of the jet. The main features of the KE are described in \citep{Qiu05,Itoh10}.
The Kuroshio and the KE transport a large amount of heat and release that to the atmosphere strongly
affecting climate. It is a region with one of the most intense air-sea heat exchange and the highest eddy kinetic energy
level. It is also a region with commercial fishing grounds of Pacific saury, tuna, squid,
Japanese sardine and other species.

The Kuroshio--Oyashio frontal zone contains various types of mesoscale and submesoscale eddies that transfer heat, salt, nutrients, carbon,
pollutants and other tracers across the ocean. They originate, besides from the KE, from
the Tsugaru Warm Current, flowing between the Honshu and Hokkaido islands, and
from the cold Oyashio Current
flowing out of the Arctic along the Kamchatka Peninsula and  the Kuril Islands.
Those eddies may persist for the periods ranging from a few weeks to a couple of years and have a strong influence on
the local climate, hydrography and fishery.

A study of the role of the KE rings and their interaction with the mean flow is important by many reasons.
They act to transfer energy to the mean currents, influent on the KE jet dynamics and drive the recirculation gyres.
They transport for a long distance water masses with biophysical properties different from ambient waters that
may have a great impact on living organisms. The strongest mesoscale eddies of both polarities are generated along the KE.
The warm-core anticyclonic rings (ACR) are pinched off from the meandering KE
mainly to the north whereas the cold-core
cyclonic ones (CR)~--- to the south of it.
The occurrence, distribution and behavior of the anticyclonic and cyclonic KE rings, moving
generally westward due to the planetary $\beta$-effect, have been studied in a number of papers
via hydrographic observations,  infrared imaging
and altimetry data \citep{Tomosada86,Ebuchi01,Waseda03,Itoh10}. However,
the process of their separation from the parent jet is not fully understood.

Lagrangian tools have been successfully used to obtain a detailed description of different advective transport phenomena
in the ocean and atmosphere. There is a vast literature on this topic (for a review see
\citep{Mancho04,Wiggins05,KP06,P13} and references therein).
As to the problem of eddy separation from strong jet currents and a cross-jet transport, there are papers  on
Lagrangian approach to the Loop Current eddy separation in the Gulf of Mexico~\citep{Kuznetsov02,Andrade13} and on
Lagrangian description of cross-jet transport in the Kuroshio Current~\citep{Mancho10,Mancho12}.
In Refs.~\citep{Kuznetsov02,Andrade13}
near-surface velocity fields from numerical models of circulation in the Gulf of Mexico have been used to study
the eddy separation  process by computing effective invariant manifolds~\citep{Kuznetsov02}  and
finite-time Lyapunov exponents \citep{Andrade13}. It has been shown there that the Lagrangian methods are
a useful supplement to traditional approaches as they reveal flow details
not easily extracted from Eulerian point of view.

In Refs.~\citep{Mancho10,Mancho12} a special lobe technique from
dynamical systems theory~\citep{Wiggins:1992:CTDS} and a method of distinguished hyperbolic trajectories
~\citep{Ide02}
have been applied to find a geometrical skeleton of some transport processes in an altimetric velocity field in
the Kuroshio region including surface cross-jet transport.
In Refs.~\citep{Samelson,UlBP07,JETP10,PRE10}
an detailed analysis, revealing mechanisms of chaotic zonal and cross-jet transport,
has been carried out for a few kinematic and dynamical analytic models of
meandering jets.

In this paper we study numerically the process
of interaction of cold-core cyclonic rings with the KE main
current, the events of their separation from the parent jet and their role
in near-surface cross-jet transport.
The special aim is to know whether it was possible for
Fukushima-derived radionuclides to cross the KE jet which is supposed to be
an impenetrable barrier.
Simulation is based on solving advection equations for synthetic particles
in the AVISO velocity fields.
The results are plotted as 1)~backward-in-time Lagrangian latitudinal maps,
where colors code
the latitudes from which particles in a given region came to their final
positions on the map
and 2)~tracking maps showing where they were walking and how frequently
visited different places in the region for a given period of time.
In Sec.~3.1 we compute both types of the maps to trace origin
of water masses inside two CRs pinched off from the KE jet in summer 2011 and to document the surface cross-jet
transport. The Fukushima derived Cs isotopes are used
as Lagrangian tracers to study
transport and mixing processes. We apply the material line technique
in Sec.~3.2 to trace the origin of
water parcels with measured levels of concentrations of Fukushima derived Cs isotopes collected
in two R/V cruises in June and July 2011 in the large area of the Northwest Pacific~\citep{Kaeriyama13,Buesseler12}.
The results of the simulation are supported by tracks of the surface drifters
which were deployed in the area.

\section{Data and methodology}
\begin{figure}[!htb]
\centerline{\includegraphics[width=0.5\textwidth]{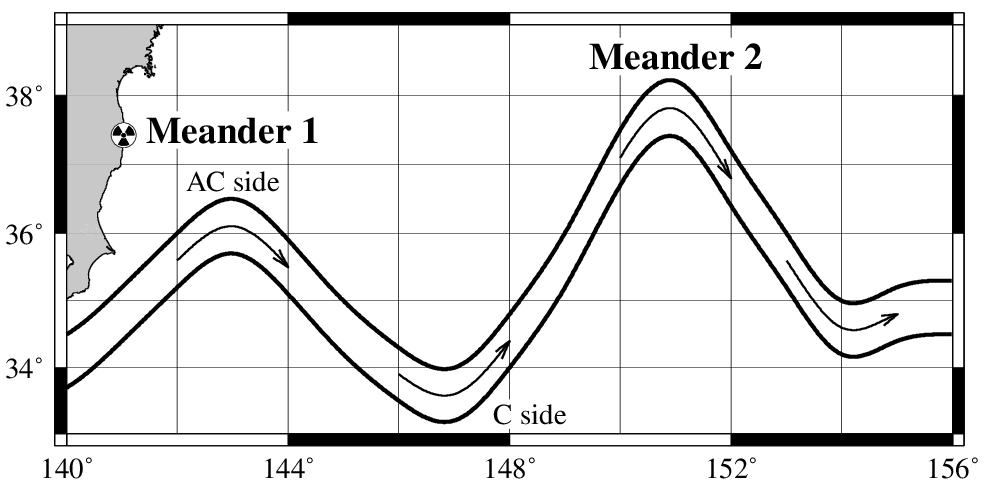}}
\caption{Schematic view of the Kuroshio Extension state with two quasistationary meanders.
Location of the FNPP is shown.}
\label{fig1}
\end{figure}
Geostrophic velocities  were obtained from the AVISO database (\url{http://www.aviso.oceanobs.com}).
The data is gridded on a $1/3^{\circ}\times 1/3^{\circ}$ Mercator grid.
Bicubical spatial interpolation and third order Lagrangian polynomials in time are used to provide
accurate numerical results. Lagrangian trajectories are computed by integrating the advection equations with a fourth-order Runge-Kutta scheme
with a fixed time step of $0.001$th part of a day.
\begin{figure*}[!htb]
\centerline{\includegraphics[width=0.5\textwidth]{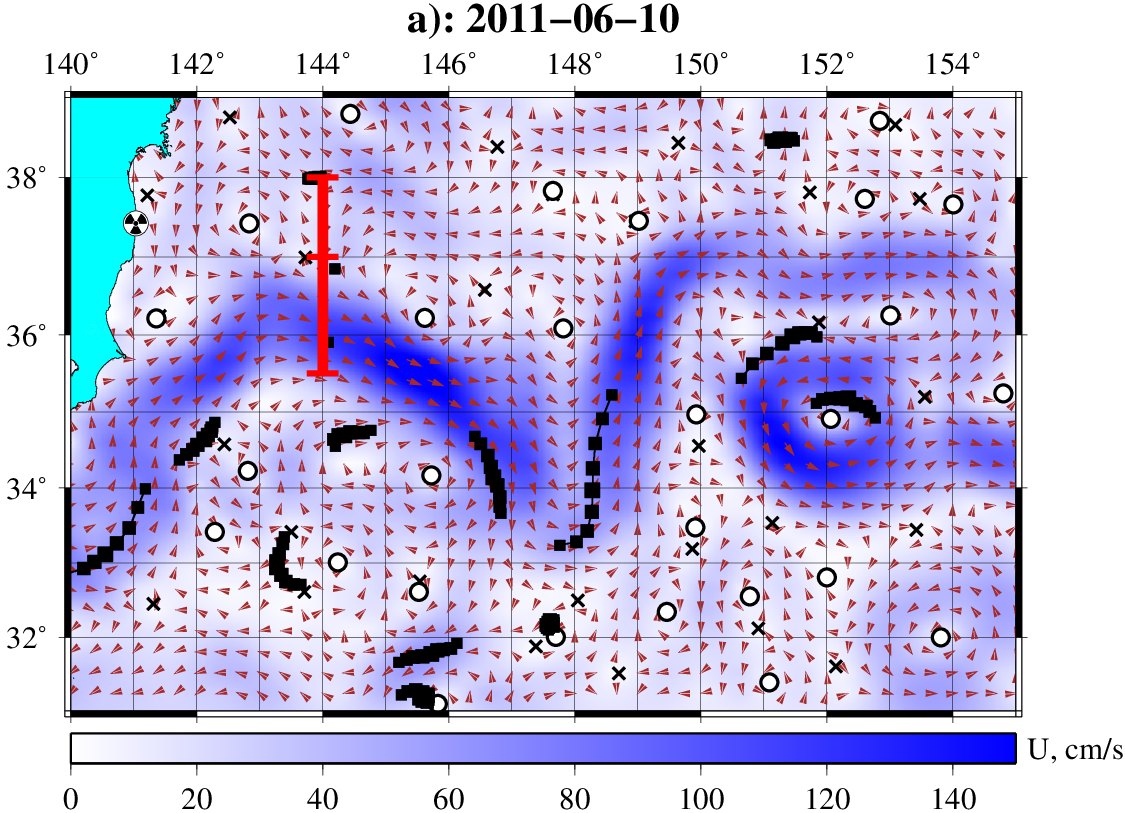}
\hfill
\includegraphics[width=0.5\textwidth]{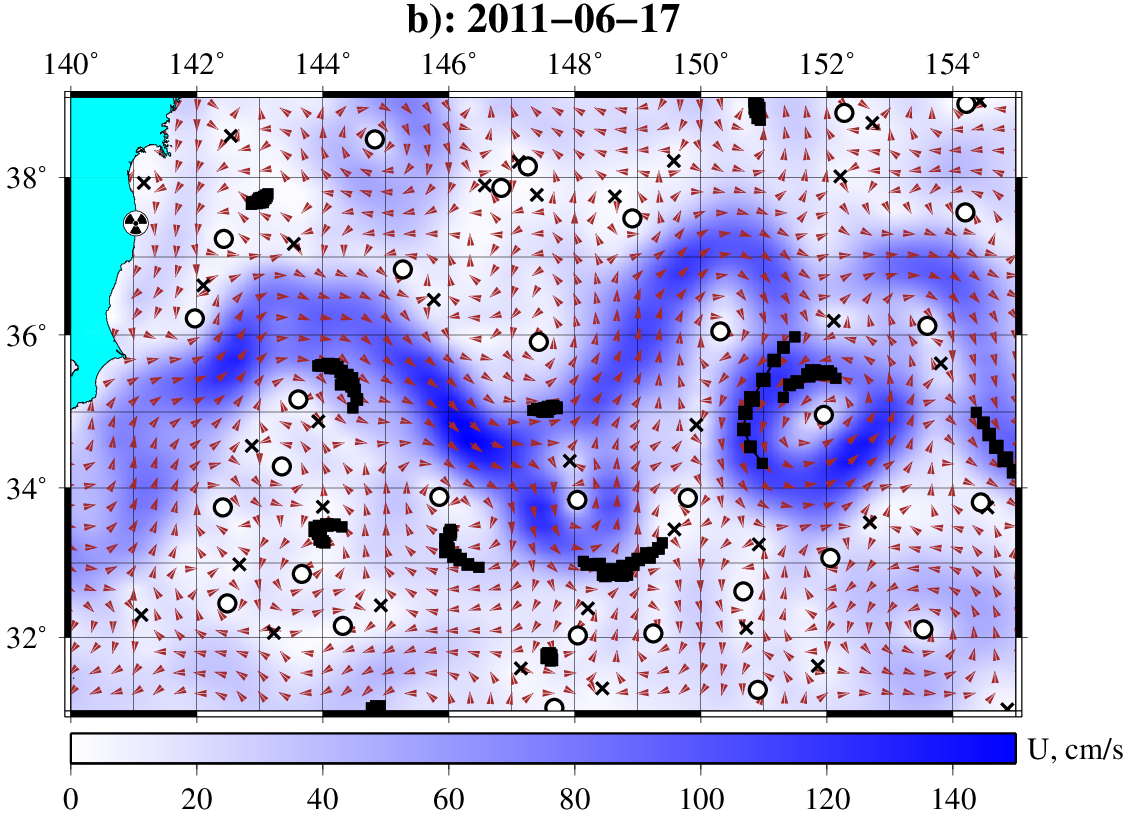}}
\centerline{$\mathstrut$}
\centerline{\includegraphics[width=0.5\textwidth]{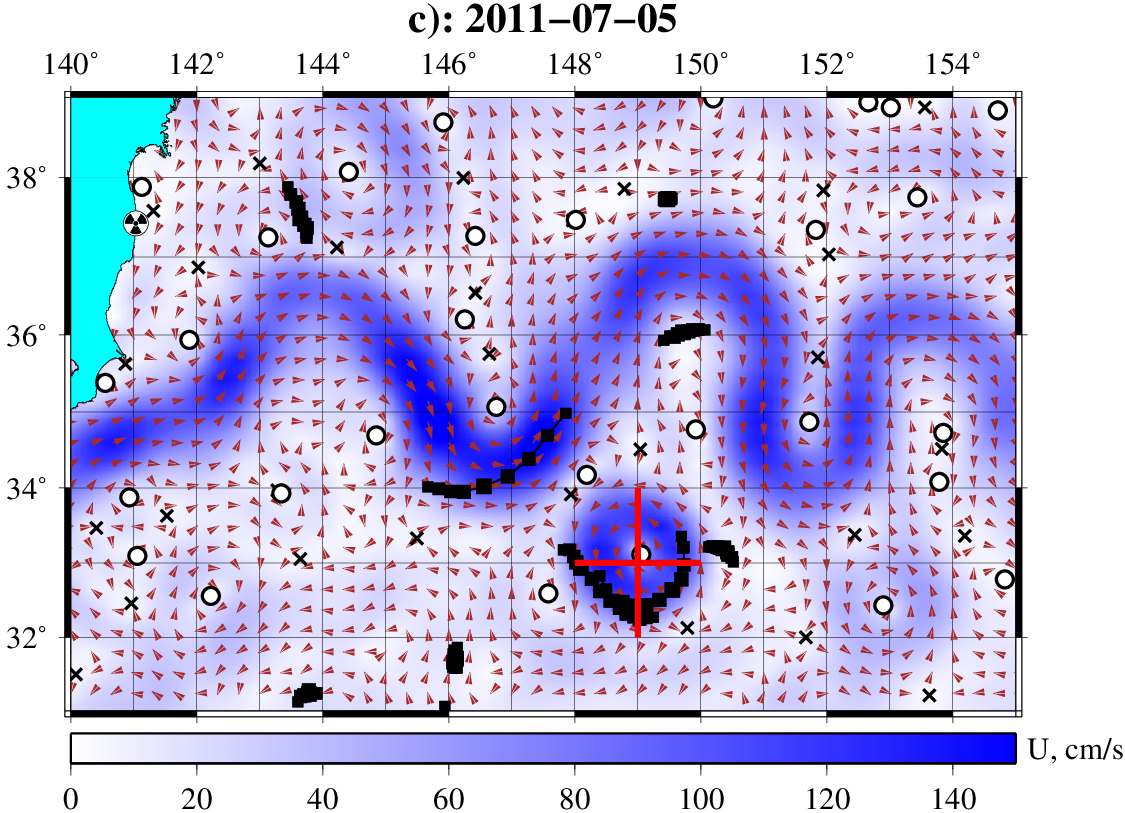}
\hfill
\includegraphics[width=0.5\textwidth]{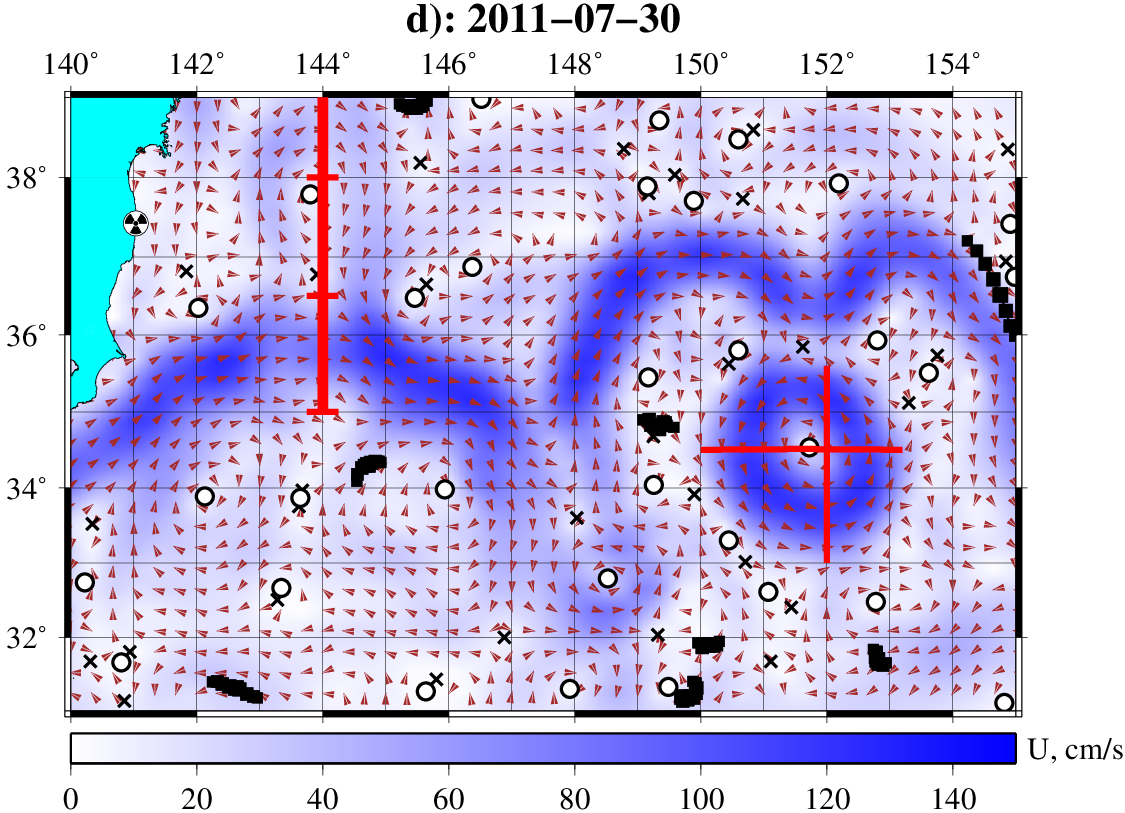}}
\caption{a), b) and c) Metamorphoses of the velocity field in the process of formation of the CR1 on 10 June, 17 June
and 5 July 2011, respectively, with tracks of drifters imposed.
d) Velocity field on 30 July 2011 with the CR2 separated from the jet.
The red straights are material lines that have been evolved backward in time
to trace origin of the corresponding markers.}
\label{fig2}
\end{figure*}
\begin{figure}[!htb]
\centerline{\includegraphics[width=0.5\textwidth]{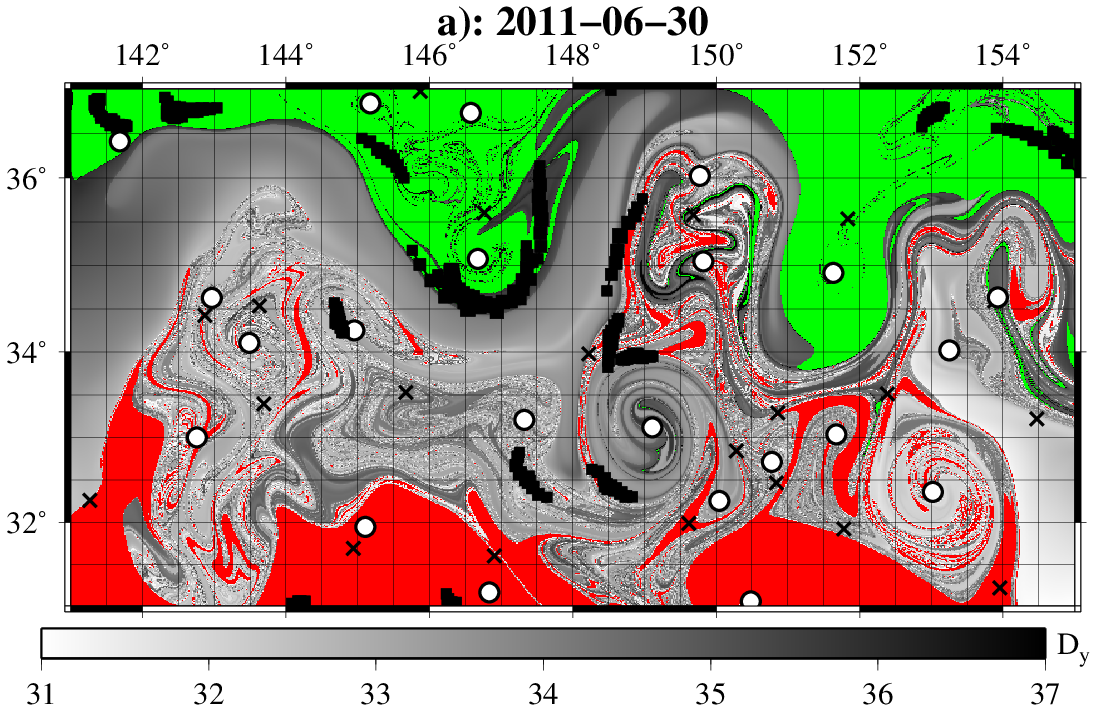}}
\centerline{\includegraphics[width=0.5\textwidth]{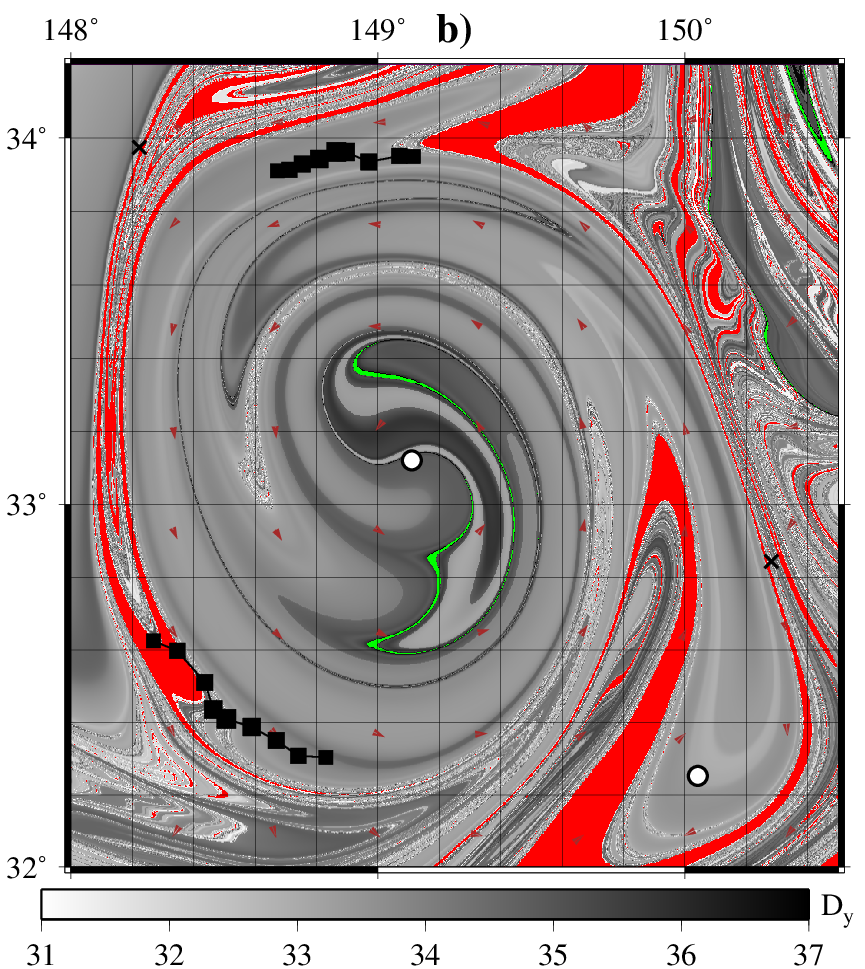}}
\caption{a) Lagrangian latitudinal map computed from
30 June to 11 March 2011 with ``red'' and ``green'' particles originated
from the latitudes ${<}31^{\circ}$~N and ${>}37^{\circ}$~N,
respectively. Nuances of the grey color in geographic degrees code the
particles originated from the latitudes between $31^{\circ}$~N and $37^{\circ}$~N.
b) Zoom of the area with the CR1 demonstrates transport across the KE jet (``green'' particles).
Tracks of two drifters to be trapped by the CR1 are shown as black squares.}
\label{fig3}
\end{figure}

Satellite altimetry demonstrates clearly that the KE alternates between two dominant states:
one with two quasistationary meanders and another one when the meanders are not especially prominent
\citep{Qiu05,Sugimoto12}. During the meandering state steep troughs develop stimulating ring pinchoff events on the both flanks of the
jet. In this paper we will focus on that state with increased eddy activity.  A sketch of the meandering KE state
in  Fig.~\ref{fig1} shows the eastward jet current with two crests
near $x=143^{\circ}$~E and $x=151^{\circ}$~E which are anticyclonic side of the jet and two troughs
near $x=147^{\circ}$~E and $x=153^{\circ}$~E with cyclonic rotations. In reality the KE jet is highly unstable,
the meander's amplitude may change in the course of time, and locations of the crests and troughs may
fluctuate strongly both in meridional and zonal directions.

Motion of a fluid particle in a two-dimensional flow is the trajectory
of a dynamical system with  given initial conditions governed by the velocity
field. The corresponding advection equations are written as follows:
\begin{equation}
\frac{d x}{d t}= u(x,y,t),\quad \frac{d y}{d t}= v(x,y,t),
\label{adveq}
\end{equation}
where the longitude, $x$, and the latitude, $y$, of a passive particle
are in geographical minutes, $u$ and $v$ are angular zonal
and meridional components of the velocity expressed in minutes per day.

The Lagrangian technique we use is based on calculation of particle motion in an altimetric velocity field
forward and backward in time. When integrating advection equations (\ref{adveq}) forward in time we compute
particle trajectories to know their fate and  when integrating them backward in time we know origin of particles
and their history. A graphic view of transport and mixing in a studied area is provided by so-called
Lagrangian synoptic maps which are plots of one of the Lagrangian indicators versus particle's initial positions
\citep{OM11,P13,FAO13}. The region under study is seeded by a large number of synthetic particles whose
trajectories are computed forward or backward in time for a given period of time.
There are a number of Lagrangian indicators (or descriptors in terminology of \citep{Mancho12}) that may be used
to characterize transport and mixing processes in the ocean and atmosphere.
Among them are  zonal and absolute displacements of water parcels, the number of their cyclonic and
anticyclonic rotations, vorticity, their residence time in a given area, the number of times particles visit
different places in the region and others. The so-called M-function, which measures
the Euclidean arc-length of the curves outlined by trajectories for a finite-time interval \citep{Mancho09}, can be also
used to plot the corresponding Lagrangian maps. Sometimes it is useful to compute
composite maps with two or more Lagrangian indicators plotted together.
To compute the Lagrangian maps we apply the methodology elaborated in our recent papers and applied to study
transport and mixing processes in different basins, from marine bays \citep{FAO13} and seas
\citep{OM11} to the ocean scale \citep{DAN11,P13}.

The material line technique used here is a powerful tool to trace origin, history and fate of water masses.
A large number of synthetic particles (markers) are placed on a meridional and/or zonal line, crossing a feature
under study, and evolve forward or backward in time. This method is especially useful if one
place material lines along the transects with in-situ measurements.
In Sec.~3.2 we carry out such simulations with
material lines placed along the transects in the Northwest Pacific where concentrations of radioactive
Cs isotopes have been measured in surface seawater during two R/V cruises
in 2011 after the FNPP accident.
Our simulations are complimented by tracks of available surface
drifters (\url{http://www.aoml.noaa.gov/phod/dac/index.php}).

\section{Results}
\subsection{Lagrangian study of origin of Kuroshio Extension rings
and possibility of surface cross-jet transport}

Documenting separation of rings from jets, their merging with jets
and tracking their propagation are
longstanding problems in oceanography.
Altimeter velocity field along with drifter observations
may be used with these aims.
Near surface velocity fields on fixed days are shown in  Fig.~\ref{fig2} where crosses and open circles mark
``instantaneous'' hyperbolic and elliptic stagnation points, respectively. Color codes the value of the
velocity $U$ in cm/s. Tracks of some drifters for 3 days are plotted by black squares.

In  Fig.~\ref{fig2} we illustrate the process of formation of a CR (which we denote as CR1)
from the KE meander in June 2011 when the KE was in the state with two prominent
quasistationary meanders. In the beginning of June the trough
of the first meander starts to steepen with its edges becoming day by day
closer and closer to each other (Fig.~\ref{fig2}a on 10 June).
The edges merge eventually closing a volume of water with a cyclonic
rotation to be connected with the parent jet by an arch (Fig.~\ref{fig2}b on 17 June).
An elliptic point in its center and a hyperbolic point in the neck of the meander appear.
Approximately five days later, the velocity field bifurcates, the ring CR1
is separated from the jet and the meander
amplitude decreases correspondingly (Fig.~\ref{fig2}c on 5 July).
Tracks of two drifters, encircling partly the ring,
are shown in Fig.~\ref{fig2}c.
\begin{figure*}[!htb]
\centerline{\includegraphics[width=0.5\textwidth]{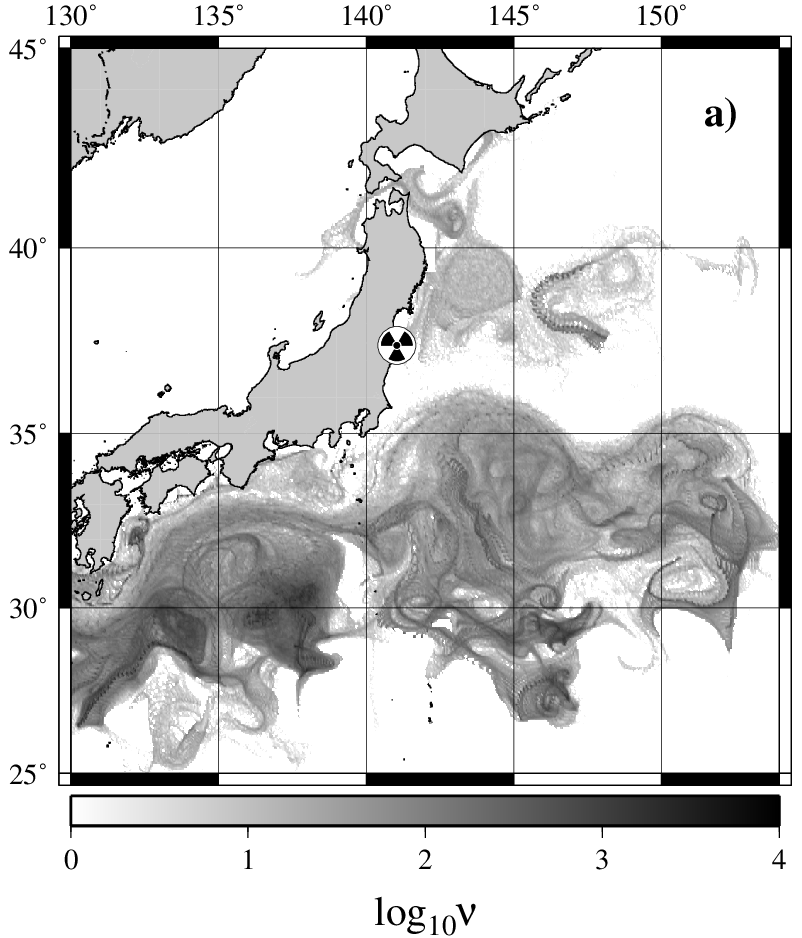}
\hfill
\includegraphics[width=0.5\textwidth]{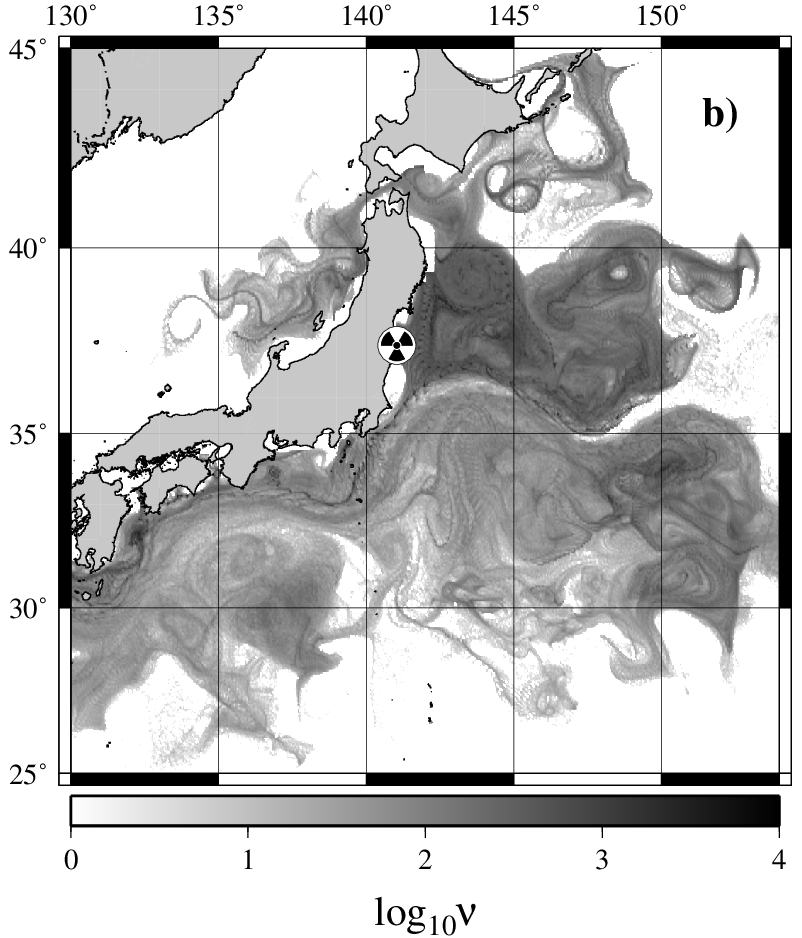}}
\caption{Tracking maps for the markers placed
on material lines crossing a) CR1 (see Fig.~\ref{fig2}c) and b) CR2
(see Fig.~\ref{fig2}d). The maps show where the corresponding
markers were walking after the FNPP accident, from 11 March to 10 April 2011. The density of traces is in the logarithmic scale.}
\label{fig4}
\end{figure*}

Transport of water masses across strong jet currents,
like the Gulf Stream and the Kuroshio, is important because
they separate waters with distinct bio-physico-chemical properties.
It may cause heating and freshing of waters with a great impact on
the weather and living organisms.
As to transport across the KE, one should mention the paper~\citep{Mancho10}
where
a turnstile mechanism \citep{Wiggins:1992:CTDS} has been supposed to provide KE cross-jet
transport in altimetric data sets.
\citep{Mancho10} have computed stable and unstable manifolds of
relevant distinguished
hyperbolic trajectories and found a lobe transport across a Lagrangian barrier
defined from pieces of those manifolds. It is a mechanism of chaotic advection
well studied with analytical model flows \citep{Wiggins:1992:CTDS,KP06}.
However, it is not a direct evidence of cross-jet transport because of
difficulties and inevitable errors in computing the manifolds and
relevant trajectories in altimetric data sets.
Moreover, the cross-jet transport has been found by ~\citep{Mancho10}
in a far downstream region
between $155^{\circ}$~E and
$165^{\circ}$~E where the KE jet is highly unstable and may even bifurcate.

The question whether the much more stable upstream KE jet between $141^{\circ}$~E and $153^{\circ}$~E
is an impenetrable barrier for cross-jet transport is still open.
The new aspect of that problem arised suddenly after the Fukushima accident.
By the common opinion~\citep{Buesseler12}, it is difficult to expect
observation of Fukushima-derived radionuclides on the southern side
of the KE jet. Could contaminant waters from the Fukushima area cross
the KE jet and appear on the southern side of the jet or not?

In order to document directly the cross-jet transport of Fukushima-derived radionuclides,
we compute a special kind of Lagrangian maps
by integrating the advection equations
(\ref{adveq}) for a large number of particles in the study area
backward in time from a fixed day till the day of the accident (11 March 2011)
and calculating their meridional displacements $D_y$.
Colors on such maps code the latitudes from which the corresponding particles
came to their final positions on the map. The latitudinal map on 30 June 2011
in Fig.~\ref{fig3}a demonstrates that ``red'' waters crossed the latitude
$31^{\circ}$~N from the south, ``green'' waters crossed the latitude of the FNPP
($141^{\circ}05'$~E, $37^{\circ}25'$~N) from the north whereas nuances of
the grey color
code the particles originated from the latitudes between $31^{\circ}$~N and $37^{\circ}$~N.
Zoom in Fig.~\ref{fig3}b shows the CR1 with ``green''
water in its core (originated from the latitudes ${>}37^{\circ}$~N) that may contain
increased concentration of Fukushima-derived radionuclides.

The simulation results are
confirmed by tracks of two surface drifters which were trapped by the CR1
(Fig.~\ref{fig3}b).
The southern track belongs to
the drifter released in the beginning of January 2011. It was transported by
the Kuroshio Current from the southwest. The northern track belongs to
the drifter No.~36473 released on 11 June in
the R/V ``{\it Ka'imikai-o-Kanaloa}'' cruise~\citep{Buesseler12} at the point
${\sim} 144^{\circ}$~E and ${\sim} 36^{\circ}$~N. It crossed the KE jet
and was trapped by the CR1. Its track is also visible at the periphery of
the CR1 in Fig.~\ref{fig2}c. It is a ``green'' drifter in Fig.~2
in Ref.~\citep{Buesseler12}. Some of ``red'' drifters in that Fig.~2,
deployed in the cruise, have been trapped by the meander trough that
formed the CR2 in July 2011.

The map in Fig.~\ref{fig3}b  is a clear evidence of transport of water
across the KE jet. To check that finding
we initialized a material line crossing the CR1 core along
$31^{\circ}06'$~N (the horizontal red line in Fig.~\ref{fig2}c)
and evolved it backward in time till the
day of the accident. It has been shown that the fragments of that line,
containing ``green'' particles, really came from the area nearby the location of
the FNPP  whereas the other particles have been advected
to form the CR1 from the west, mainly along the KE jet.
However, the amount of potentially dangerous Fukushima waters in the core
of the CR1 is comparatively small.

Now we apply the material line technique to trace the origin of water masses in
the CR1 and in a large CR named as CR2
that was born after a separation of the trough of the second KE meander
from the parent jet in July 2011. In June--July 2011 it was deforming
strongly and eventually produced a ring-like structure with a diameter
of about 300 km that has been detached to the south from the parent jet
and then reabsorbed in a short time. That ring-like structure with the center
near $x=34^{\circ}30'$~N and $y=152^{\circ}$~E
is seen in Fig.~\ref{fig2}d in the altimetric velocity field on 30 July 2011.

Starting on 30 June, we evolve backward in time two perpendicular material lines with a large number of
markers, crossing the CR1 (Fig.~\ref{fig2}c). The other two
material lines were chosen to cross the CR2 on 30 July (Fig.~\ref{fig2}d)
when a hyperbolic point appeared between the ring and the KE jet.
Coming back to the question, whether the KE jet is an impenetrable barrier
for Fukushima-derived radionuclides,
we compute backward-in-time tracking maps. The region under study is
divided in a large number of small cells, and
one fixes how many times markers visited each cell during the month after the accident
when the maximal leakage directly into the ocean
and atmospheric fallout on the ocean surface have been registered.
The result for CR1 markers is shown in Fig.~\ref{fig4}a where the
density of marker's traces,
$\nu$, is in the logarithmic scale. First of all,
the probability that the CR1 contains the contaminant water is
comparatively small because the density of points in the region around the
FNPP is small.
This map confirms the result of direct calculation of the cross-jet transport
in Fig.~\ref{fig3}b where only a small amount of potentially contaminant
``green'' water, originated from the latitudes ${>}37^{\circ}$~N, is visible.
It is clear from Fig.~\ref{fig4}a that the CR1 consists mainly of
Kuroshio water.

As to CR2 markers, the density of their traces in the area,
that is supposed to be contaminated, is much higher as compared
to the CR1 case (Fig.~\ref{fig4}b). It means that the probability to observe
higher concentrations of
Fukushima-derived radionuclides in surface waters of the CR2 is expected to be comparatively
large. The CR2 contains water parcels that have moved during the month after
the accident around the mesoscale eddies to be present
to the north and east from the FNPP location.
To the day of the accident there was the eddy system with
a large anticyclonic warm-core Kuroshio ring (ACR) with the center around
$144^{\circ}$~E and $39^{\circ}$~N, a small anticyclonic eddy to the north
of it and a medium
cyclonic eddy at the traverse of the Tsugaru Strait. It has been shown in our
paper~\citep{DAN11} that namely this eddy system has governed mixing and transport of
radioactive water part of which has been trapped by the
ACR and advected around the adjacent eddies to the north whereas the very
ACR moved slowly to the south.
The concentration of radionuclides around those eddies might be significantly
greater than in other places because they are a kind of attractors~\citep{DAN11}.
The influence of the ACR is evident on both the tracking maps as
a patch with increased density of traces at its place.

We conclude this section by emphasizing that the material line technique may
be useful to finding the surface areas
in the North Pacific that are potentially dangerous due to the risk of
radioactive contamination. Before choosing
the track of a planed R/V cruise, it is instructive to make a simulation by
initializing backward-in-time evolution of
material lines, crossing eddies in the
region visible in the velocity field and on Lagrangian maps.
The tracking maps computed in such experiments would help to know where
one could expect higher or lower concentrations of Fukushima-derived
radionuclides in this or that eddy.
\begin{figure*}[!hpt]
\centerline{\includegraphics[width=0.49\textwidth]{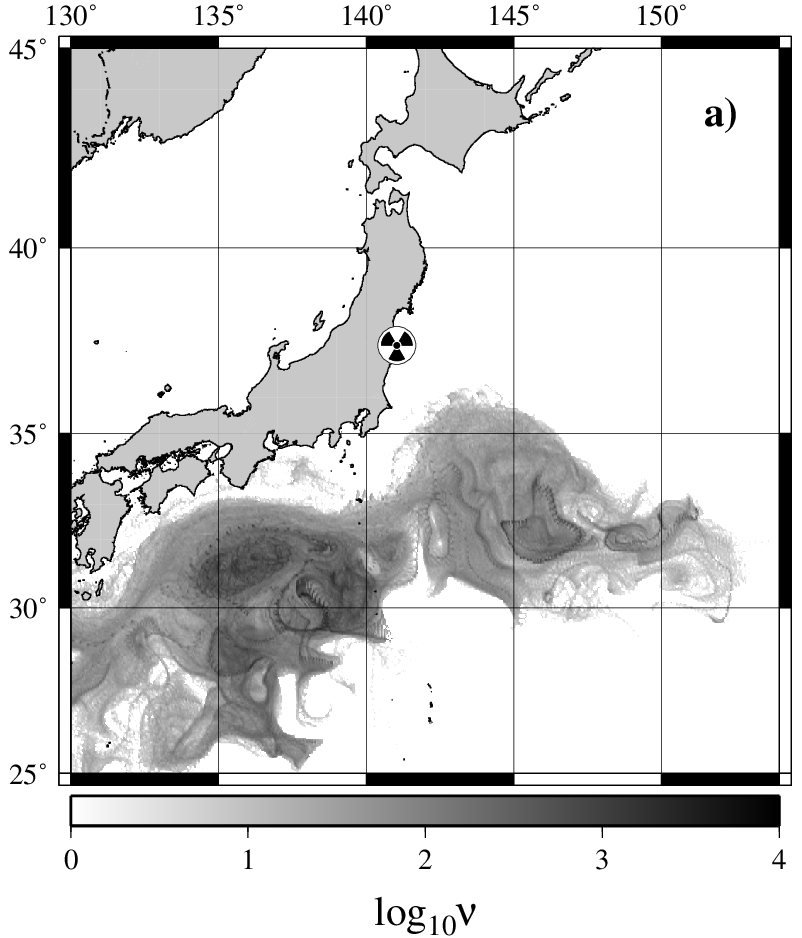}
\hfill
\includegraphics[width=0.49\textwidth]{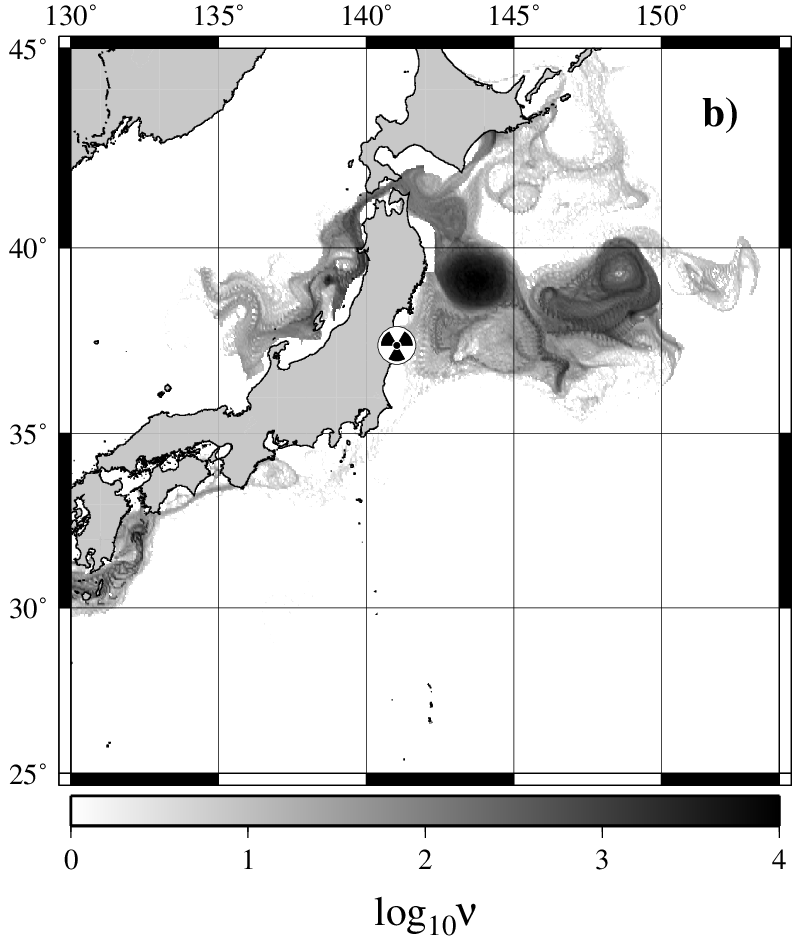}}
\centerline{$\mathstrut$}
\centerline{\includegraphics[width=0.49\textwidth]{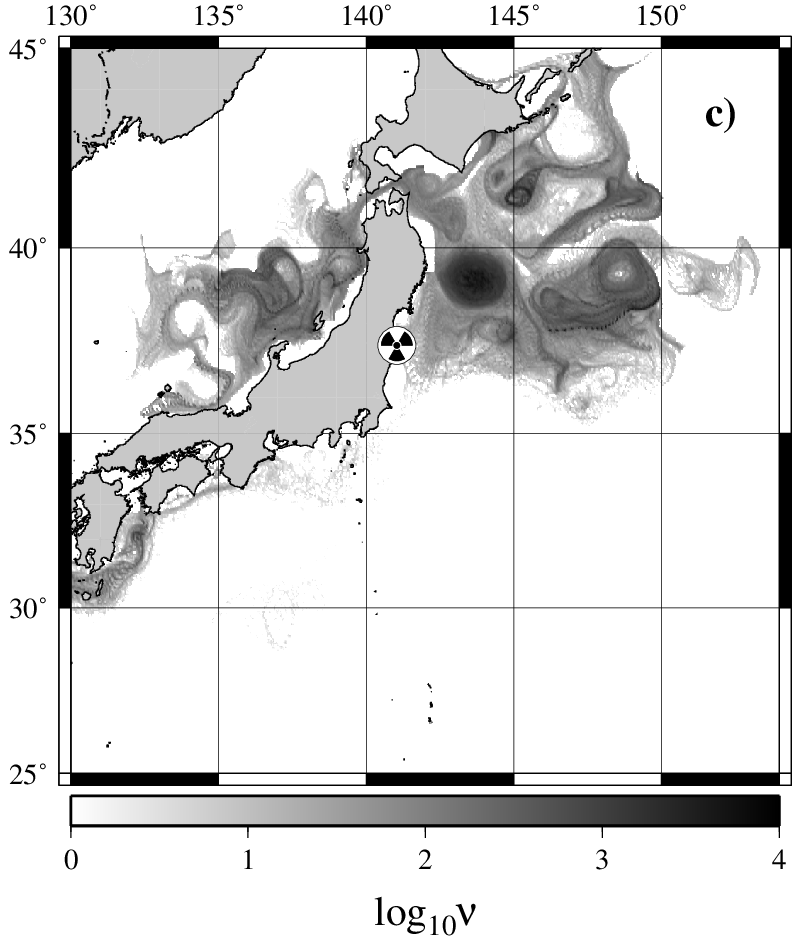}
\hfill
\includegraphics[width=0.49\textwidth]{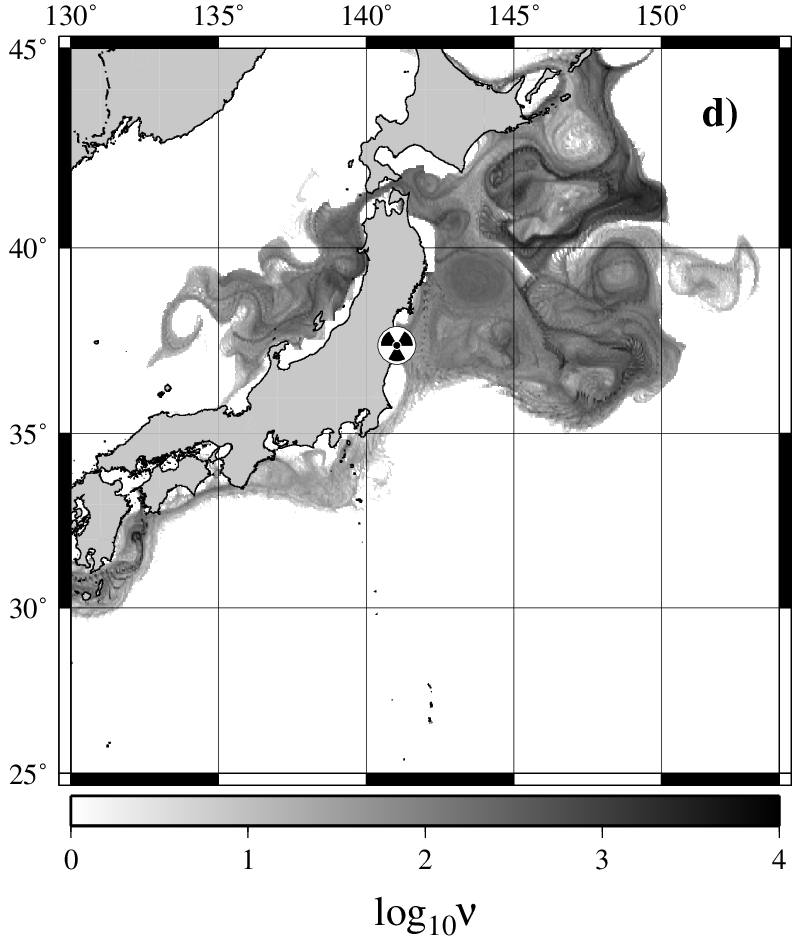}}
\caption{Tracking maps for the markers placed
on four material line segments (see Fig.~\ref{fig2}d)
along the transect where seawater samples
have been collected in 26--29 July 2011 in the R/V ``{\it Kaiun maru}''
cruise~\citep{Kaeriyama13}. The maps show where the corresponding
markers were walking from 11 March to 10 April 2011.}
\label{fig5}
\end{figure*}
\begin{figure*}[ht]
\centerline{\includegraphics[width=0.5\textwidth]{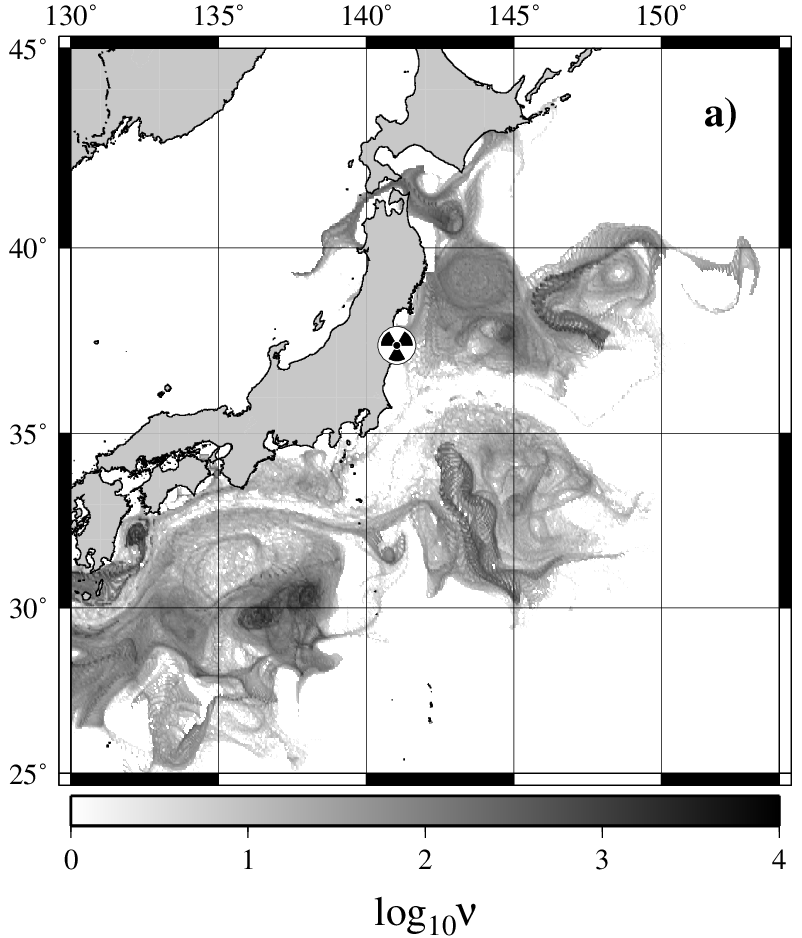}
\hfill
\includegraphics[width=0.5\textwidth]{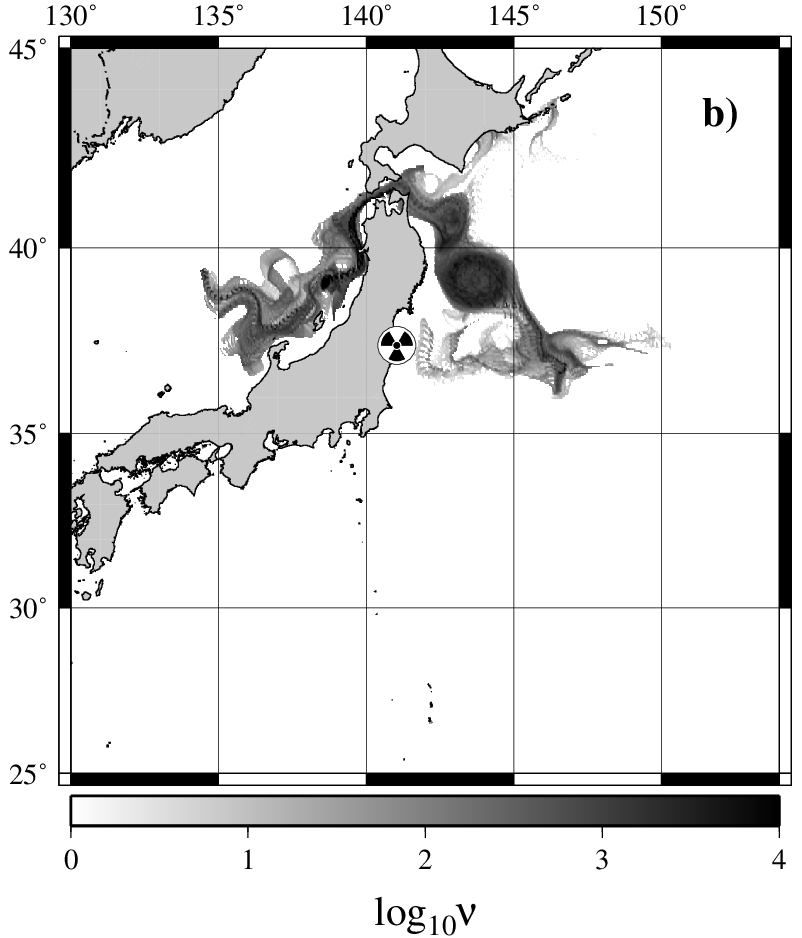}}
\caption{The same as in Fig.~\ref{fig5} but with markers placed
on two material line segments (see Fig.~\ref{fig2}a)
along the transect where seawater samples
have been collected in 10 and 11 June 2011 in the R/V ``{\it Ka'imikai-o-Kanaloa}''
cruise~\citep{Buesseler12}.}
\label{fig6}
\end{figure*}

\subsection{Fukushima-derived radionuclides as Lagrangian tracers}

In this section we apply the material line technique to trace the origin of
water parcels with measured levels of $^{134}$Cs and $^{137}$Cs
concentrations  collected
in two R/V cruises  in June and July 2011 \citep{Kaeriyama13,Buesseler12}.
Starting from the dates of sampling, we evolve backward in time material lines placed along the transects,
where stations with collected surface water samples were located.
Results of direct observation of radioactive Cs in surface seawater collected from R/V ``{\it Kaiun maru}'' in
a broad area in
the western and central North Pacific in July, October 2011 and July 2012 have been reported in Ref.~\citep{Kaeriyama13}.
In this study, we focus on the results of measurements to be carried out at the stations C43--C55
from 26 to 29 July 2011 along the $144^{\circ}$~E meridian from  $35^{\circ}$~N to $41^{\circ}$~N. That transect is
partly shown in Fig.~\ref{fig2}d. Its southern edge crosses the crest of the first KE meander whereas the northern
edge crosses partly the ACR which is visible in the
altimetric velocity field in Fig.~\ref{fig2}d with an elliptic point
in its center near  $x=144^{\circ}$~E and $y=38^{\circ}$~N. It is the same
ACR that has moved to the south from March to August and was mentioned in the end of the preceding section.
The measured $^{137}$Cs
concentrations at the stations C43--C55 have been varied from the background level of $1.9\pm 0.4$~mBq~kg$^{-1}$
(station C52) to $153\pm 6.8$~mBq~kg$^{-1}$ (station C47). The ratio $^{134}\text{Cs}/^{137}\text{Cs}$ was close to 1.
The level of the concentration of the caesium isotopes in the sea surface waters in the North Pacific before
the accident did not exceed 2--3~mBq~kg$^{-1}$.

We placed a material line along that transect and divided it into the four segments:
1)~the segment~1, $35^{\circ}$--$36^{\circ}30'$~N, crossing the first meander's crest with initialization on 26 July,
2)~the segment~2, $36^{\circ}30'$--$38^{\circ}$~N, crossing the southern edge of the ACR with initialization
on 27 July, 3)~the segment~3, $38^{\circ}$--$39^{\circ}30'$~N, crossing the core of the ACR with initialization
on 28 July and 4)~the segment~4, $39^{\circ}30'$--$41^{\circ}$~N, crossing the northern edge of the ACR with initialization
on 29 July (not shown in Fig.~\ref{fig2}d).

The tracking maps  in Fig.~\ref{fig5} shows where markers of the corresponding segments
were walking from 11 March to 10 April 2011. Markers from the segment 1,
as expected, were advected to their places in the end of July mainly by the
Kuroshio Current from the southwest and did not cross the latitude of the segment's northern end, $36^{\circ}30'$~N,
(Fig.~\ref{fig5}a). The risk of their radioactive contamination is small. It is confirmed by
the measured $^{137}$Cs concentrations at the stations C52--C55 in that segment to be 2--5~mBq~kg$^{-1}$~\citep{Kaeriyama13}
that is slightly higher than the background level.
Markers from the second segment
have been found walking mainly in the area to the north from the latitude $36^{\circ}30'$~N (Fig.~\ref{fig5}b).
Part of the initial material line crossed the ACR. That is why we see increased density of points
at the place of that ring in March and April.
A comparatively high level of Fukushima-derived caesium isotopes
is expected in the corresponding water samples. It is really the case. The $^{137}$Cs concentrations
at the stations C49 and C50 of that segment were measured to be $36 \pm 3.3$ and $50 \pm 3.6$~mBq~kg$^{-1}$~\citep{Kaeriyama13}.
The caesium concentration levels up to $153 \pm 6.8$~mBq~kg$^{-1}$~\citep{Kaeriyama13}
have been measured at the stations C46, C47 and C48 (our third segment) and
C43, C44 and C45 (our fourth segment). The tracking maps in Fig.~\ref{fig5}c and d show clearly an
increased density of traces of the corresponding markers in the area where the maximal leakage directly into the ocean
and atmospheric fallout on the ocean surface have been registered from 11 March to 10 April 2011. Those maps
also demonstrate strong mixing in the Kuroshio--Oyashio frontal zone. Markers, to be initialized at
the material lines with the length of $1.5^{\circ}$, left traces in the area $25^{\circ}\times 25^{\circ}$,
including some parts of the Sea of Japan and the Okhotsk Sea.

Fukushima-derived $^{134}$Cs and $^{137}$Cs were measured in surface and subsurface waters,
as well as in zooplankton and fish, at 50 stations in 4--18 June 2011 during
the R/V ``{\it Ka'imikai-o-Kanaloa}'' cruise (see ref.~\citep{Buesseler12} and a supplement to that paper).
We initialize a material line as shown in Fig.~\ref{fig2}a where $^{137}$Cs concentrations have been
measured on 10 and 11 June at 25 stations in the range
from $1.4\pm 0.2$~mBq~kg$^{-1}$ (station 13) to $173.6\pm 9.9$~mBq~kg$^{-1}$ (station 10).
The ratio $^{134}\text{Cs}/^{137}\text{Cs}$ was close to~1. Some markers are placed on the southern segment,
$35^{\circ}30'$--$36^{\circ}30'$~N, crossing the first meander's crest (11 June), and
the other ones~--- on the northern segment, $37^{\circ}$--$38^{\circ}$~N, to the north from the meander's crest
(10 June).
Traces of the markers of the southern segment are found on the both sides of the KE jet, whereas
traces of the markers of the northern one are on the northern side of the jet
only.
The lower part of the southern segment crosses the jet itself but its
upper part is outside of it (Fig.~\ref{fig2}a). That is why the tracking map in Fig.~\ref{fig6}a consists of
two disconnected domains, one is to the south of the jet and another one is to the north.
The measured $^{137}$Cs concentrations at the stations 13 and 14, situated in the southern segment,
were at the background level, in the range 1.4--3.6~mBq~kg$^{-1}$~\citep{Buesseler12}, because the corresponding
markers were advected by the KE current (Fig.~\ref{fig6}a).

Density of traces of the markers from the northern  segment is
comparatively high in the area around the FNPP  (Fig.~\ref{fig6}b). This finding is confirmed by
measurements at stations 10, 11 and 12 \citep{Buesseler12} where
the concentrations of Fukushima-derived $^{134}$Cs and $^{137}$Cs were in the range
21.9--173.6~mBq~kg$^{-1}$. The northern segment crosses partly the ACR visible in Fig.~\ref{fig2}a, and
the traces of its markers are dense at the place of that ring.
Some markers were advected to their places on the initial segment by the Tsushima Current from the
Sea of Japan to the Pacific Ocean through the Tsugaru Strait.

\section{Summary}

In this research, we used numerical simulations to study near-surface large-scale transport
in the KE area based on AVISO altimetric velocity field.
After solving advection equations for passive markers backward in time, we have computed Lagrangian
maps for their displacements and tracking maps for the number of times markers visited different places in the region.

Two KE cold-core cyclonic rings, CR1 and CR2, have been chosen to
illustrate the process of pinching off from the main jet in summer 2011.
The tracking and Lagrangian maps were computed to trace the origin of water
masses in the cores of those rings. They revealed near-surface cross-jet transport.
This conclusion is supported by tracks of the surface drifters
which were deployed in the area.
Water masses, constituting the CR1, have been advected mainly from the southwest by the Kuroshio and the KE,
and only a small
amount of its water was originated from the area to the north from the KE jet.
Traces of the CR2 markers have been found in a large area to the south
and the north from the jet, including the area around the FNPP location.
It is interesting that through a year, in June 2012, we have found a cold-core CR to be separated from the jet
approximately at the same place as the CR1 did in June 2011. It was confirmed by the track of a surface drifter
circulating around that CR. Simulation showed that it contained Fukushima derived markers
which were able to cross the KE jet.

We used Fukushima derived caesium isotopes as Lagrangian tracers comparing the results of our simulations
with in-situ observations of $^{134}$Cs and $^{137}$Cs
concentrations in water samples collected in two R/V cruises in June and July 2011 \citep{Kaeriyama13,Buesseler12}.
Evolving backward in time material lines along the transects where measurements have been carried out in those cruises,
we computed the corresponding tracking maps.  It is shown that
the water parcels with caesium concentration, exceeding greatly the background level, were walking during one month
after the accident in the area where the maximal leakage directly into the ocean
and atmospheric fallout on the ocean surface have been registered. The density of traces of markers with low caesium
concentration in that area was found to be comparatively small.

We would like to emphasize that the tracking technique elaborated in this paper may
be useful to planning R/V cruises in the ocean. Before choosing
the track of a planed R/V cruise, it is instructive to make a simulation by
initializing backward-in-time evolution of
material lines, crossing potentially interesting coherent structures in the
region visible in the velocity field and on Lagrangian maps.
The corresponding tracking maps would help to know where
one could expect higher or lower concentrations of radionuclides, pollutants or other Lagrangian
tracers.

\section*{Acknowledgments}
This work was supported  by the Russian Foundation for Basic Research
projects 11--01--12057, 12--05--00452, 13--05--00099 and 13-01-12404). The altimeter products were distributed by AVISO with support from CNES.

\bibliographystyle{copernicus}
\bibliography{paper}

\begin{thebibliography}{26}
\providecommand{\natexlab}[1]{#1}
\providecommand{\url}[1]{{\tt #1}}
\providecommand{\urlprefix}{URL }
\expandafter\ifx\csname urlstyle\endcsname\relax
  \providecommand{\doi}[1]{doi:\discretionary{}{}{}#1}\else
  \providecommand{\doi}{doi:\discretionary{}{}{}\begingroup
  \urlstyle{rm}\Url}\fi

\bibitem[{Andrade-Canto et~al.(2013)Andrade-Canto, Sheinbaum, and {Zavala
  Sans\'{o}n}}]{Andrade13}
Andrade-Canto, F., Sheinbaum, J., and {Zavala Sans\'{o}n}, L.: {A Lagrangian
  approach to the Loop Current eddy separation}, Nonlinear Processes in
  Geophysics, 20, 85--96, \doi{10.5194/npg-20-85-2013}, 2013.

\bibitem[{Buesseler et~al.(2012)Buesseler, Jayne, Fisher, Rypina, Baumann,
  Baumann, Breier, Douglass, George, Macdonald, Miyamoto, Nishikawa, Pike, and
  Yoshida}]{Buesseler12}
Buesseler, K.~O., Jayne, S.~R., Fisher, N.~S., Rypina, I.~I., Baumann, H.,
  Baumann, Z., Breier, C.~F., Douglass, E.~M., George, J., Macdonald, A.~M.,
  Miyamoto, H., Nishikawa, J., Pike, S.~M., and Yoshida, S.: {Fukushima-derived
  radionuclides in the ocean and biota off Japan}, Proceedings of the National
  Academy of Sciences, 109, 5984--5988, \doi{10.1073/pnas.1120794109}, 2012.

\bibitem[{Ebuchi and Hanawa(2001)}]{Ebuchi01}
Ebuchi, N. and Hanawa, K.: {Trajectory of Mesoscale Eddies in the Kuroshio
  Recirculation Region}, Journal of Oceanography, 57, 471--480,
  \doi{10.1023/A:1021293822277}, 2001.

\bibitem[{Ide et~al.(2002)Ide, Small, and Wiggins}]{Ide02}
Ide, K., Small, D., and Wiggins, S.: {Distinguished hyperbolic trajectories in
  time-dependent fluid flows: analytical and computational approach for
  velocity fields defined as data sets}, Nonlinear Processes in Geophysics, 9,
  237--263, \doi{10.5194/npg-9-237-2002}, 2002.

\bibitem[{Itoh and Yasuda(2010)}]{Itoh10}
Itoh, S. and Yasuda, I.: {Characteristics of Mesoscale Eddies in the
  Kuroshio--Oyashio Extension Region Detected from the Distribution of the Sea
  Surface Height Anomaly}, Journal of Physical Oceanography, 40, 1018--1034,
  \doi{10.1175/2009JPO4265.1}, 2010.

\bibitem[{Kaeriyama et~al.(2013)Kaeriyama, Ambe, Shimizu, Fujimoto, Ono,
  Yonezaki, Kato, Matsunaga, Minami, Nakatsuka, and Watanabe}]{Kaeriyama13}
Kaeriyama, H., Ambe, D., Shimizu, Y., Fujimoto, K., Ono, T., Yonezaki, S.,
  Kato, Y., Matsunaga, H., Minami, H., Nakatsuka, S., and Watanabe, T.: {Direct
  observation of ${}^{134}$Cs and ${}^{137}$Cs in surface seawater in the
  western and central North Pacific after the Fukushima Dai-ichi nuclear power
  plant accident}, Biogeosciences Discussions, 10, 1993--2012,
  \doi{10.5194/bgd-10-1993-2013}, 2013.

\bibitem[{Koshel' and Prants(2006)}]{KP06}
Koshel', K.~V. and Prants, S.~V.: {Chaotic advection in the ocean},
  Physics-Uspekhi, 49, 1151--1178, \doi{10.1070/PU2006v049n11ABEH006066}, 2006.

\bibitem[{Kuznetsov et~al.(2002)Kuznetsov, Toner, Kirwan~Jr., Jones, Kantha,
  and Choi}]{Kuznetsov02}
Kuznetsov, L., Toner, M., Kirwan~Jr., A., Jones, C., Kantha, L., and Choi, J.:
  {The Loop Current and adjacent rings delineated by Lagrangian analysis of the
  near-surface flow}, Journal of Marine Research, 60, 405--429, 2002.

\bibitem[{Madrid and Mancho(2009)}]{Mancho09}
Madrid, J. A.~J. and Mancho, A.~M.: {Distinguished trajectories in time
  dependent vector fields}, Chaos: An Interdisciplinary Journal of Nonlinear
  Science, 19, 013\,111, \doi{10.1063/1.3056050}, 2009.

\bibitem[{Mancho et~al.(2004)Mancho, Small, and Wiggins}]{Mancho04}
Mancho, A.~M., Small, D., and Wiggins, S.: {Computation of hyperbolic
  trajectories and their stable and unstable manifolds for oceanographic flows
  represented as data sets}, Nonlinear Processes in Geophysics, 11, 17--33,
  \doi{10.5194/npg-11-17-2004}, 2004.

\bibitem[{Mendoza and Mancho(2012)}]{Mancho12}
Mendoza, C. and Mancho, A.~M.: {The Lagrangian description of aperiodic flows:
  a case study of the Kuroshio Current}, Nonlinear Processes in Geophysics, 19,
  449--472, \doi{10.5194/npg-19-449-2012}, 2012.

\bibitem[{Mendoza et~al.(2010)Mendoza, Mancho, and Rio}]{Mancho10}
Mendoza, C., Mancho, A.~M., and Rio, M.-H.: {The turnstile mechanism across the
  Kuroshio current: analysis of dynamics in altimeter velocity fields},
  Nonlinear Processes in Geophysics, 17, 103--111,
  \doi{10.5194/npg-17-103-2010}, 2010.

\bibitem[{Prants et~al.(2011{\natexlab{a}})Prants, Budyansky, Ponomarev, and
  Uleysky}]{OM11}
Prants, S., Budyansky, M., Ponomarev, V., and Uleysky, M.: {Lagrangian study of
  transport and mixing in a mesoscale eddy street}, Ocean Modelling, 38,
  114--125, \doi{10.1016/j.ocemod.2011.02.008}, 2011{\natexlab{a}}.

\bibitem[{Prants(2013)}]{P13}
Prants, S.~V.: {Dynamical systems theory methods to study mixing and transport
  in the ocean}, Physica Scripta, 87, 038\,115, \doi{10.1088/0031-8949}, 2013.

\bibitem[{Prants et~al.(2011{\natexlab{b}})Prants, Uleysky, and
  Budyansky}]{DAN11}
Prants, S.~V., Uleysky, M.~Y., and Budyansky, M.~V.: {Numerical simulation of
  propagation of radioactive pollution in the ocean from the Fukushima Dai-ichi
  nuclear power plant}, Doklady Earth Sciences, 439, 1179--1182,
  \doi{10.1134/S1028334X11080277}, 2011{\natexlab{b}}.

\bibitem[{Prants et~al.(2013)Prants, Ponomarev, Budyansky, Uleysky, and
  Fayman}]{FAO13}
Prants, S.~V., Ponomarev, V.~I., Budyansky, M.~V., Uleysky, M.~Y., and Fayman,
  P.~A.: {Lagrangian analysis of mixing and transport of water masses in the
  marine bays}, Izvestiya, Atmospheric and Oceanic Physics, 49, 82--96,
  \doi{10.1134/S0001433813010088}, 2013.

\bibitem[{Qiu and Chen(2005)}]{Qiu05}
Qiu, B. and Chen, S.: {Variability of the Kuroshio Extension Jet, Recirculation
  Gyre, and Mesoscale Eddies on Decadal Time Scales}, Journal of Physical
  Oceanography, 35, 2090--2103, \doi{10.1175/JPO2807.1}, 2005.

\bibitem[{Samelson and Wiggins(2006)}]{Samelson}
Samelson, R.~M. and Wiggins, S.: {Lagrangian Transport in Geophysical Jets and
  Waves: The Dynamical Systems Approach (Interdisciplinary Applied
  Mathematics)}, Springer, 2006.

\bibitem[{Sugimoto and Hanawa(2012)}]{Sugimoto12}
Sugimoto, S. and Hanawa, K.: {Relationship between the path of the Kuroshio in
  the south of Japan and the path of the Kuroshio Extension in the east},
  Journal of Oceanography, 68, 219--225, \doi{10.1007/s10872-011-0089-1}, 2012.

\bibitem[{Tomosada(1986)}]{Tomosada86}
Tomosada, A.: {Generation and decay of Kuroshio warm-core rings}, Deep Sea
  Research Part A. Oceanographic Research Papers, 33, 1475--1486,
  \doi{10.1016/0198-0149(86)90063-4}, 1986.

\bibitem[{Uleysky et~al.(2007)Uleysky, Budyansky, and Prants}]{UlBP07}
Uleysky, M.~Y., Budyansky, M.~V., and Prants, S.~V.: {Effect of dynamical traps
  on chaotic transport in a meandering jet flow}, Chaos: An Interdisciplinary
  Journal of Nonlinear Science, 17, 043105, \doi{10.1063/1.2783258}, 2007.

\bibitem[{Uleysky et~al.(2010{\natexlab{a}})Uleysky, Budyansky, and
  Prants}]{JETP10}
Uleysky, M.~Y., Budyansky, M.~V., and Prants, S.~V.: {Chaotic transport across
  two-dimensional jet streams}, Journal of Experimental and Theoretical
  Physics, 111, 1039--1049, \doi{10.1134/S1063776110120174},
  2010{\natexlab{a}}.

\bibitem[{Uleysky et~al.(2010{\natexlab{b}})Uleysky, Budyansky, and
  Prants}]{PRE10}
Uleysky, M.~Y., Budyansky, M.~V., and Prants, S.~V.: {Mechanism of destruction
  of transport barriers in geophysical jets with Rossby waves}, Physical Review
  E, 81, 017\,202, \doi{10.1103/PhysRevE.81.017202}, 2010{\natexlab{b}}.

\bibitem[{Waseda(2003)}]{Waseda03}
Waseda, T.: {On the eddy-Kuroshio interaction: Meander formation process},
  Journal of Geophysical Research, 108, \doi{10.1029/2002JC001583}, 2003.

\bibitem[{Wiggins(1992)}]{Wiggins:1992:CTDS}
Wiggins, S.: {Chaotic Transport in Dynamical Systems}, {Interdisciplinary
  applied mathematics}, Springer-Verlag, 1992.

\bibitem[{Wiggins(2005)}]{Wiggins05}
Wiggins, S.: {The dynamical systems approach to {L}agrangian transport in
  oceanic flows}, Annual Review of Fluid Mechanics, 37, 295--328,
  \doi{10.1146/annurev.fluid.37.061903.175815}, 2005.

\end{thebibliography}

\end{document}